\documentclass[11pt]{article}

\usepackage[left=2.3cm,right=2.3cm,top=2.9cm,bottom=2.9cm]{geometry} 

\usepackage{color}
\usepackage{graphicx}
\usepackage{amsmath}
\usepackage{amssymb}
\usepackage{fancyhdr}

\usepackage[sort&compress]{natbib}
\usepackage[colorlinks=true,citecolor=blue,linkcolor=black]{hyperref}

\renewenvironment{abstract}{%
\hspace*{1cm}\begin{minipage}{0.87\textwidth}
}
{\end{minipage}}
\makeatletter
\renewcommand\@maketitle{%
\hfill
}

\begin{document}

\pagestyle{empty}

\vspace*{2cm}

To appear in the \textit{Journal of Coupled Systems and Multiscale Dynamics} \\[0.5ex]
American Scientific Publishers, 2015\\[0.5ex]
doi:10.1166/jcsmd.2015.1069

\vspace*{2cm}

{\LARGE \bf Data-driven modelling of biological multi-scale processes}
\\[3ex]
{\large \it Jan Hasenauer\,$^{1,2,*}$, Nick Jagiella\,$^{1}$, Sabrina Hross\,$^{1,2}$ and Fabian J. Theis\,$^{1,2}$}
\\[3ex]
{\it
$^{1}$Helmholtz Zentrum M\"unchen - German Research Center for Environmental Health, \\
\hspace*{1mm} Institute of Computational Biology, 85764 Neuherberg, Germany \\[1ex]
$^{2}$Technische Universit\"at M\"unchen, Center for Mathematics, \\
\hspace*{1mm} Chair of Mathematical Modeling of Biological Systems, 85748 Garching, Germany \\[3ex]
$^*$ Corresponding author. Email: jan.hasenauer@helmholtz-muenchen.de
}

\vspace*{1cm} 

\begin{abstract}
\noindent Biological processes involve a variety of spatial and temporal scales. A holistic understanding of many biological processes therefore requires multi-scale models which capture the relevant properties on all these scales. In this manuscript we review mathematical modelling approaches used to describe the individual spatial scales and how they are integrated into holistic models. We discuss the relation between spatial and temporal scales and the implication of that on multi-scale modelling. Based upon this overview over state-of-the-art modelling approaches, we formulate key challenges in mathematical and computational modelling of biological multi-scale and multi-physics processes. In particular, we considered the availability of analysis tools for multi-scale models and model-based multi-scale data integration. We provide a compact review of methods for model-based data integration and model-based hypothesis testing. Furthermore, novel approaches and recent trends are discussed, including computation time reduction using reduced order and surrogate models, which contribute to the solution of inference problems. We conclude the manuscript by providing a few ideas for the development of tailored multi-scale inference methods.

\vspace*{2ex}\noindent\textit{\bf Keywords}: 
Systems Biology, Computational Biology, Multi-scale Modelling, Parameter Estimation, Model Selection, Surrogate Models.
\end{abstract}

\vspace*{0.5cm}

\newpage
\setcounter{tocdepth}{2}
\tableofcontents

\pagestyle{fancy}

\section{Introduction}
The key aim of mathematical, computational and systems biology is to achieve a holistic understanding of biological systems~\cite{Kitano2002}. Over decades this aim was mainly approached by moving from the study of single molecules to the analysis of biological networks. Starting with the groundbreaking work of Hodgkin \& Huxley~\cite{HodgkinHux1952}, who described the dynamics of individual neurones using ordinary differential equations, a successful revolution began, which resulted in a significantly improved understanding of gene regulation, signalling, metabolism and many other processes~\cite{KlippBook2005}. Large-scale models for individual pathways and collections of pathways have been developed (see e.g.~\cite{SchoeberlEic2002,DuarteHer2004,Klipp2005,AlbeckBur2008b,Schlatter2009b,XuVys2010,BachmannRau2011}) along with powerful theoretical and computational methods. 

Despite great successes it has been recognised that the study of cellular networks is only the first step, as multi-cellular organisms are more than a collection of pathways. A mechanistic understanding of complex biological functions and processes requires the consideration of multiple spatial and temporal scales. During the last two decades, multi-scale models for a variety of processes have been developed such as the beating of the heart~\cite{Nobel2002,HunterBor2003}, the development of cancer~\cite{AndersonQua2008} in all its subtypes, the drug delivery and action~\cite{EissingKue2011,SchallerWil2013}, or the data processing in our brain~\cite{PetzoldAlb2008}. These multi-scale and multi-physics models were derived by expanding existing models using top-down, bottom-up or so called middle-out approaches~\cite{BruggemanWes2007}, as well as by integrating existing models for different aspects. 

As the development of coherent multi-scale models requires expertise on different biological scales, large consortia have been formed, bringing together researches to tackle complex questions. One great example is the World Physiome Initiative with the Physiome Project~\cite{HunterBor2003,PhysiomeProject}, which unites researches from all around the globe. Among others, the Physiome Project demonstrated that multi-scale models can indeed deepen our insights~\cite{tenTusscherNob2004}, e.g., of the human heart, by integrating models for different processes. Depending on the question, a significant benefit in general does not even require the consideration of all possible (intermediate) biological scales~\cite{SchallerWil2013}.

Despite the demonstrated applicability of multi-scale models, many important theoretical results and computational methods for their analysis are still missing. Multi-scale models are often obtained by coupling different model classes, e.g., logical models, ordinary differential equations (ODE), partial differential equations (PDE) or agent-based models, and a coherent theory for model properties such as stability, sensitivity and bifurcations, is mostly not present. More importantly from a practical point of view, methods for the model-based multi-scale data integration are hardly available. Model-based data integration via parameter estimation~\cite{Tarantola2005} is essential for the evaluation of the consistency of models and the underlying hypotheses with experimental findings. It enables hypothesis testing and generation, uncertainty quantification, as well as the prediction of processes dynamics under unknown conditions. Model-based data integration was maybe the most important reason why mechanistic pathway models were so successful in the past~(see \cite{SchoeberlPac2009}). To exploit the full potential of multi-scale models the collection of rigorous theoretical and computational methods has to be extended. 

The field of multi-scale modelling has grown rapidly over the last years. Recent reviews focus on the modelling and simulation \cite{HunterBor2003,MartinsFer2010,DadaMen2011,WalpolePap2013}. Methodological challenges and potential solutions, e.g.~in the context of parameter estimation for multi-scale models, are hardly addressed by the published reviews. As this discussion is in our opinion urgently necessary, we complement previous reviews by focussing on multi-scale model inference, in particular on methods for optimisation, uncertainty analysis and model selection. Furthermore, we discuss novel approaches which can help to advance inference of multi-scale models. We introduce concepts to decrease the computational complexity associate to multi-scale simulations, including reduced order and surrogate modelling. Furthermore, interesting theoretical concepts for the analysis of interconnected systems are highlighted.

The review is structured as follows: In Section~\ref{sec: modelling} common single- and multi-scale modelling approaches are summarised from a mathematical perspective. Furthermore, dependencies between spatial and temporal scales are discussed, along with a few groundbreaking contributions to the field of multi-scale modelling. In Section~\ref{sec: analysis and simulation}, we outline available analysis tools for multi-scale models, discuss the reasons for the limitations and provide suggestions for future research approaches. In Section~\ref{sec: parameter estimation and model selection} -- the central part of this paper -- we turn our attention to the data-driven modelling of multi-scale biological processes. We outline available optimisation and uncertainty analysis methods for model-based multi-scale data integration, discuss the limitations arising from computational complexity and outline methods which could be used to overcome them. The manuscript concludes in Section~\ref{sec: conclusion}.

\section{Modelling}
\label{sec: modelling}
The term ``multi-scale models'' is widely used but for many years the scientific community lacked an appropriate definition. One has recently been provided by the \textit{Inter\-agency Modelling and Analysis Group (IMAG)}~\cite{IMAG2015}:
\\[2ex]
\textit{``Multi-scale, biomedical modelling uses mathematics and computation to represent and simulate a physiological system at more than one biological scale. Biological scales include atomic, molecular, molecular complexes, sub-cellular, cellular, multi-cell systems, tissue, organ, multi-organ systems, organism, population, and behaviour. These multi-scale biomedical models may also include dynamical processes which span multiple time and length scales.''}
\\[2ex]
This definition is, like many others, very broad, in order to capture a complete research field. In the following, we will try to provide some more details on modelling approaches for individual biological scales and their integration into multi-scale models. Furthermore, we will shortly discuss the relation of time and length scales.

\subsection{Modelling of individual biological scales and processes}
In bio- and life-sciences, many different multi-scale modelling approaches are used. This variety of multi-scale modelling approaches is caused by the variety of modelling approaches for the individual biological scales (see~Figure~\ref{fig: models for individual scales} and discussion below). In the following, we outline the most common modelling approaches and model classes used for the study of individual spatial scales. Our understanding of biological scales coincides with the IMAG definition for multi-scale, biomedical modelling but to ensure stringency some scales are lumped together. For the different modelling approaches, we will outline key mathematical properties. Furthermore, we summarize how these properties are exploited to capture the processes occurring in a particular scale.

\subsubsection{Atomic and molecular scale}
The building blocks of biological systems are deoxyribonucleic acids (DNAs), ribonucleic acids (RNAs), proteins, lipids, and metabolites. The structure, formation (e.g., folding) and interaction of these biomolecules is commonly described and analysed using quantum mechanic (QM) models and molecular mechanic (MM) models~\cite{Cramer2004}. Quantum mechanic models provide the most detailed description of atomic and subatomic processes. For their simulation \textit{ab~initio}, density function and (semi-)empirical methods are available~\cite{Cramer2004}. As these methods are computationally demanding, even on very short time-scales, only small molecules can be analysed using fully quantum mechanic models. To simulate larger biomolecules, molecular mechanic models are used. These models exploit a cruder description of molecules and describe atoms as point masses and point charges which interact via spring-like interactions and van der Waals forces~\cite{BurkertAll1982}. Mathematically, molecular mechanistic models can be viewed as systems of agents which interact and these dynamics are captured by differential equations. To capture important details, e.g., in reaction centre of an enzyme, while being able to simulate large molecules, quantum mechanic / molecular mechanic (QM/MM) models have been developed~\cite{WarshelLev1976}. To enable the study of larger biomolecules, coarse-grained and lattice models are employed. These models represent groups of atoms by a point mass and charge~\cite{SmithHal2001}.

\subsubsection{Sub-cellular and cellular scale}
The interaction of biomolecules gives rise to biological pathways and processes, such as gene regulation, signal transduction and metabolism. These processes have been modelled using for example continuous-time discrete-state Markov chains (CTMCs)~\cite{Gillespie1977,McAdamArk1997,Gillespie2000}, stochastic differential equations (SDEs)~\cite{Gillespie2000}, ordinary differential equations (ODEs)~\cite{KlippBook2005}, partial differential equations (PDEs)~\cite{SchaffFin1997,MoraruLoe2005}, agent-based models~\cite{TurnerSch2004}, stochastic or deterministic Boolean models~\cite{KauffmanPet2003,KlamtHau2009,KrumsiekMar2011}, Petri nets~\cite{Chaouiya2007,MarwanWag2011} and constraint-based models~\cite{DuarteHer2004}. Boolean and Petri net models are often referred to as qualitative models, as other model classes enable a more quantitative description.

Gene regulation is mostly modelled using CTMCs~\cite{ShahrezaeiSwa2008}, Boolean models~\cite{WittmannBlo2009} and Petri nets~\cite{MarwanWag2011}. These models capture the discreteness and the stochasticity of the processes arising from the small copy number of genes. The dynamics of signalling pathways are mostly described using SDEs~\cite{Wilkinson2009,Fuchs2010} and ODEs~\cite{HodgkinHux1952,SchoeberlEic2002}, as large abundances of the involved biochemical species supports a continuous interpretation in terms of concentrations. For the assessment of the  qualitative properties of signalling pathways, also Boolean models are used~\cite{Schlatter2009b}. As the availability of spatial information increases, also PDEs, spatially resolved CTMCs and agent-based modelling approaches are more commonly used~\cite{MoraruLoe2005,KlannLap2009}. For the description of the metabolism often ODE-based formulations are exploited. Under the assumption of fast temporal dynamics, these ODE models can be used to derive constraint-based formulations which merely require information about the stoichiometry of the biochemical reactions~\cite{DuarteHer2004}.

\begin{figure*}[p]
\centering
\includegraphics[width=\textwidth]{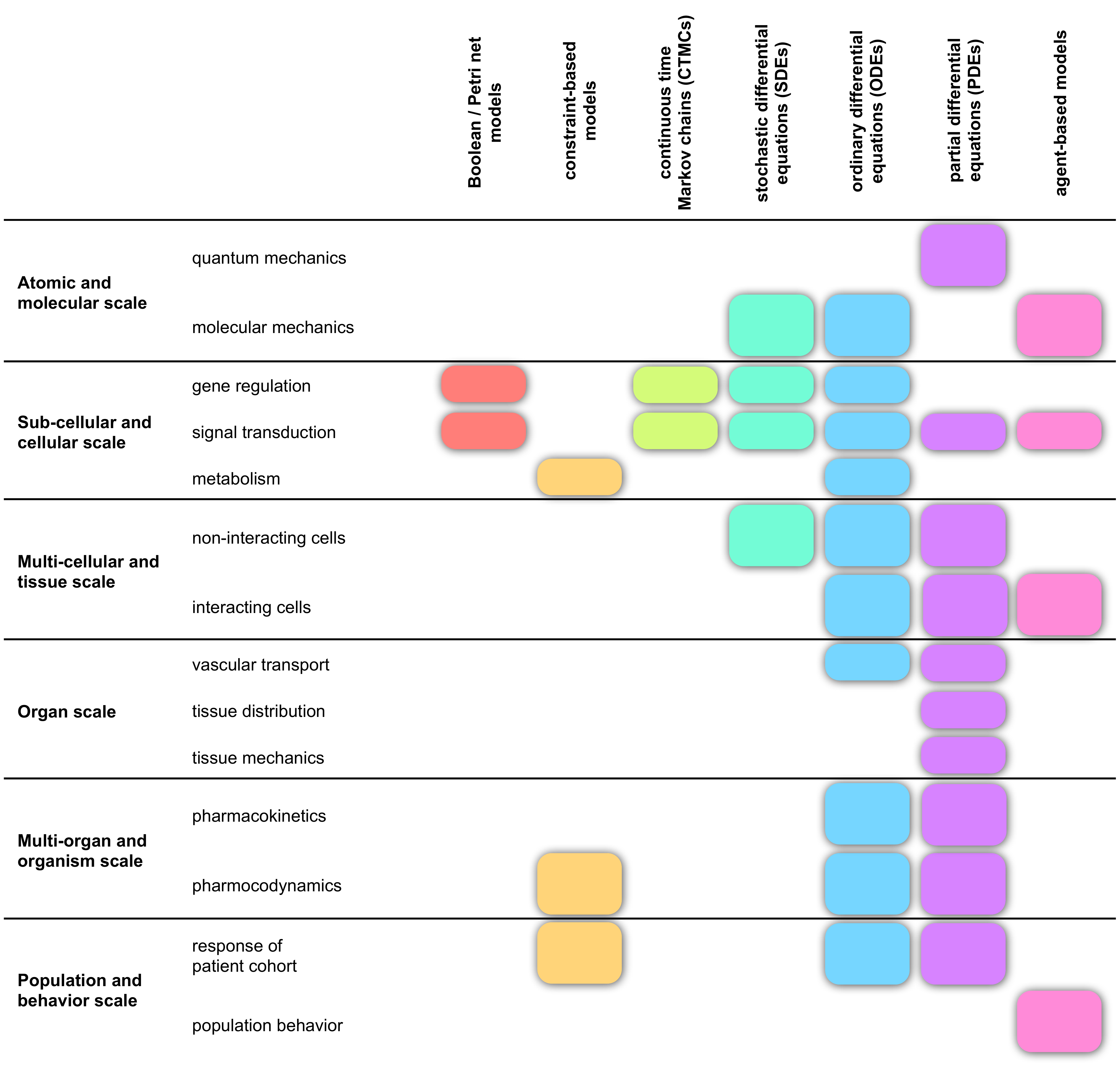}
\caption{Visual summary of the types of mathematical models used to describe biological processes on different scales. A coloured box indicates that the types of mathematical models are routinely used for models of the respective spatial scale. The spatial scales are split up in subcategories to allow for additional insight.}
\label{fig: models for individual scales}
\end{figure*}

\subsubsection{Multi-cellular and tissue scale}
The variability of individual cells and the dependence of cellular dynamics on the current cell state, the history of individual cells, environmental condition and/or the spatial location in the tissue is captured by cell population models. The first population model has been introduced by von Foerster~\cite{vonFoerster1959}. It described populations of non-interacting, proliferating cells. This groundbreaking work on structured population models -- systems of PDEs -- has been followed by theoretical contributions and model extensions (see~\cite{Trucco1965,Oldfield1966,SinkoStr1967,LuzyaninaRoo2007,BanksSut2010,HasenauerSch2012b} and references therein). To study stochastic dynamics of gene regulation and signal transduction in non-dividing, heterogeneous cell populations, the Chemical Master Equation (CME)~\cite{Gillespie1992} and the Fokker-Planck equation (FPE)~\cite{Gardiner2004} are employed. To account for mixed, stochastic and deterministic sources of cell-to-cell variability, also extensions to the Chemical Master Equation~\cite{ZechnerRue2012,StamatakisZyg2010} and the Fokker-Planck equation~\cite{HasenauerWal2011b,HasenauerPhdThesis2013} have been introduced.

The spatial organisation present in biological tissues is typically captured by using either agent-based or PDE models. The first model class uses agents to represent cells: either each cell individually (1 agent = 1 cell)~\cite{AndersonQua2008,BuskeGal2011,LangMar2011,WalkerGeo2008,Jagiella2012,HoehmeBru2010}, or coarse-grained compartments of several cells (1 agent = many cells)~\cite{Drasdo2005}, or a single cell by an ensemble of agents (many agents = 1 cell)~\cite{GranerGlazier1992,Newman2005}. The latter, e.g. cellular Potts models, enable a detailed description of cell-cell interaction and cell movement. These models allow for direct interactions of cells. An alternative to agent-based models are provided by cellular automata~\cite{GerhardtSch1989,PeakWes2004,DeutschDor2005}, which abstract the detailed biological interactions to rules. On the other hand, the second model class -- PDE models -- provide a more macroscopic view and describe the cell densities as continuum in space and capture their temporal evolution. There is significant work available on PDE models for avascular tumour growth~\cite{PreziosiTos2009,WiseLow2008}, angiogenesis~\cite{TosinAmb2006}, developmental processes~\cite{WittmannBlo2009,HockNg2013}, bacterial biofilms~\cite{EfendievBook2013} and many other important processes.

\subsubsection{Organ scale}
Organs consist of many different tissues and possess a complex spatial organisation, the length scales are however too large to model individual cells. To enable organ-scale simulation models, homogenisation methods have been developed. These methods originated in the field of porous media and use local averaging to enable PDE-based description of flow, transport and biomechanics processes in tissues as well as intracellular dynamics~\cite{HunterBor2003,BassingthwaighteRay2010,ChapmanShi2008,ErbertsederRei2012,EhlersWag2013,KarajanRoh2013}. The coupling of intra- and extracellular dynamics yields bidomain models~\cite{Whiteley2008}. Commonly, these bidomain PDE models are coupled with ODE models for the vascular transport~\cite{LiChe1993,ReicholdSta2009}. 

As organs possess different functions, their physiological characteristics are diverse. This diversity is also reflected by the models. Models of the heart~\cite{HunterBor2003,BassingthwaighteRay2010} and other muscles~\cite{KarajanRoh2013} focus on tissue mechanics while models of the lung~\cite{LiChe1993,ErbertsederRei2012}, the brain~\cite{ReicholdSta2009,EhlersWag2013}, the craniocervical region~\cite{RutkowskaHau2012}, and solid tumours~\cite{ChapmanShi2008} focus on transport and delivery processes. Accordingly, the underlying mathematical descriptions and equation types differ.

\subsubsection{Multi-organ and organism scale}
On the organism scale, the properties of and crosstalk between organs are studied, e.g., in the context of drug delivery and treatment. The crosstalk between organs is mostly mediated via the vascular and the lymphoid system, which are used for the transport of substances between organs. Additionally, the nerve systems mediates information exchange between sensory organs and muscles with the brain.

Models for the organism scale consider the body physiology~\cite{DerendorfMei1999}, e.g., the volume of organs and large blood vessels as well as the available area for substance exchange and vessel permeability. These physiological parameters, which are often well accessible, allow for the parameterisation of multi-compartment models. The individual compartments are organs and large blood vessels. In pharmacokinetics (PK), ODE-based multi-compartment models are used to describe liberation, absorption, distribution, metabolisation and excretion of drugs~\cite{RodgersLea2005,RodgersRow2006}. These pharmacokinetic models have been combined with pharmacodynamic (PD) models which describe the biochemical and physiological effects of drugs on the body~\cite{Nestorov2007,EdgintonThe2008,EdgintonWil2008}. The resulting physiology-based pharmacokinetic/pharmacodynamic (PK/PD) models~\cite{EissingKue2011} are widely used in the pharmaceutical industry for drug discovery, decision making and data integration~\cite{KuepferLip2012}.

Most PK/PD models rely on a mathematical description in terms of ODEs. To capture fast metabolic processes, recently also hybrid models have been introduced which combine ODEs and constraint-based modelling~\cite{KraussSch2012}. Furthermore, in larger consortia like the Physiome Project~\cite{PhysiomeProject} and the Virtual Liver Network~\cite{HolzhutterDra2012,KuepferKer2014}, PK/PD models are integrated with PDE and agent-based models for a refined description of the organs of interest.

\subsubsection{Population and behaviour scale}
As inter-individual variability is pervasive, personalised PK/PD models have been developed. These personalised models account, among others, for differences in gender, age, height, body mass, general health and genetic alterations~\cite{EissingKue2011}. On the population scale, this yields mixed-effect models~\cite{Pinheiro1994,WillmannHoh2007}. These mixed-effect models often rely on an ODE-based description of individuals and provide a description of the population dynamics in terms of a high-dimensional system of PDEs, a Liouville-type equation. As the dimensionality is prohibitive for a direct simulation of the PDE, representative sets of ODE simulations are used to evaluate the population statistics. The corresponding results can nowadays be exploited in personalised medicine~\cite{EissingKue2011}.

In social sciences, mathematical modelling is employed to study the interactions of individuals and the dynamics of groups~\cite{Bonabeau2002,Kennedy2012}. Therefore, agent-based models are frequently used as they provide a natural description of many processes, e.g., evacuation or traffic. Accordingly, they can be used to capture and to predict emergent phenomena.
\\[3ex]
A visualisation of the modelling approaches exploited to describe individual biological scales is provided in Figure~\ref{fig: models for individual scales}. We find that the class of modelling approaches used to capture processes on the cellular scale is most versatile. This is probably related to the focus of systems biology on the cellular scale. Furthermore, ODE and PDE models are used across scales and are the most common modelling approaches. Reason for this are most likely the existence of well established, deep theoretical results for these models~\cite{CoddingtonLev1955,Evans1998} and the availability of computationally efficient simulation methods (see~\cite{HindmarshBro2005,WendlandEfe2003} and references therein).

\subsection{Multi-scale modelling}
In the previous section various modelling approaches for biological processes have been outlined. The reason for the variety is the multi-scale and multi-physics nature of biological systems. To span different scale and processes, different models / model classes are necessary and have to be coupled. In the following, we will shortly outline different multi-scale modelling philosophies. We will discuss simulation and coupling approaches and conclude with a number of well-known multi-scale models. Throughout the chapter we will use the terms micro-scale and macro-scale to indicate small and large spatial scales respectively.

\subsubsection{Aims of multi-scale modelling}
The ultimative aim of multi-scale modelling is to describe the macroscopic behaviour of complex systems using  microscopic models derived from first-principles. Accordingly, many scientists argue that: 
\\[2ex]
\textit{``The task of multi-scale modelling is to carry out macro-scale simulations without making use of ad hoc constitutive relations. Instead the necessary micro-scale constitutive information is extracted from micro-scale models.''}~\cite{EEng2007} 
\\[2ex]
This would allow for a holistic analysis based on fully mechanistic descriptions. While the defined task is appealing, the majority of multi-scale modelling in biology follow a different philosophy.

In biology, a variety of different physics have to be covered, ranging from molecule-molecule interaction over signalling all the way to fluid and tissue mechanics. As a simultaneous first-principle modelling of these different physics is currently beyond reach, many research projects focus on the coupling of established macroscopic models to capture and study the multi-physics nature of biological systems~\cite{HunterBor2003}. Examples are multi-scale models for pharmacokinetic and pharmacodynamic models, which describe fluid transport as well as intracellular signalling~\cite{EissingKue2011,SchallerWil2013}. In a preliminary step the individual macroscopic models are derived from first-principles and than used in coupled multi-scale, multi-physics simulations. As modelling of individual processes is relatively well established, many systems and computational biologists currently aim at linking models of different processes.

The two aforementioned aims
\\[1ex]
\textbf{Aim\;1:}\;\textit{Multi-scale modelling using first-principles.}
\\[1ex]
\textbf{Aim\;2:}\;\textit{Multi-scale and multi-physics modelling using coupled macroscopic models.}
\\[1ex]
possess a significant overlap. As the approaches and methods developed to reach the different aims are however different, we will in the following discuss them separately. In Section~2.2.2 we will discuss simulation methods for multi-scale models, while in Section~2.2.3 we focus on coupled multi-scale, multi-physics models.

\subsubsection{Multi-scale modelling using first-principles}

The multi-scale modelling using first-principles relies on the simulation of macro-scale processes using models with micro-scale resolution. As this is often computationally infeasible, \textit{multi-resolution methods}~\citep{Barth2001} and \textit{equation-free modelling approaches}~\cite{KevrekidisSam2009} have been developed. Multi-resolution methods, including adaptive mesh refinement and multi-grid methods~\cite{TrottenbergOos2001,AscherHab2003}, are numerical schemes which reduce the number of micro-scale model evaluations by performing a successive (local) refinement. While multi-resolution methods are well established, the consideration of computationally demanding micro-scale models is often problematic. To circumvent long-term micro-scale simulations, equation-free modelling can be used. Equation-free modelling exploits on-the-fly coupling of microscopic and macroscopic simulations~\cite{EEng2003}. A macroscopic model/solver is used, which might require missing macroscopic information. This missing macroscopic information is obtained using short microscopic simulations. Therefore, the microscopic model is first constrained using the available macroscopic data (=~lifting), under these constraints the microscopic model is evaluated for a short time interval (=~microscopic simulator), and the resulting simulation data are used to constrain the macroscopic model (=~restriction). Several equation-free modelling approaches have been proposed, among others, \textit{heterogeneous multi-scale methods}~\cite{EEng2003}, \textit{(coarse) projective integration} and the \textit{gap-tooth scheme}~\cite{KevrekidisGea2003}. All these methods treat microscopic models as black boxes and merely use simulation results, which led to the term equation-free modelling.

\subsubsection{Multi-scale and multi-physics modelling using coupled macroscopic models}
\label{sec: Multi-physics modelling using coupled macroscopic models}
While the available mathematical results and computational approaches for the development of first-principles multi-scale models are promising, application examples in biology and medicine are limited. In biology the modelling of complex multi-physics systems is so far mostly approached by coupling models which describe different aspects of the underlying biological systems.

A nice example is the integrated physiologically-based whole-body model recently published by Schaller et al.~\cite{SchallerWil2013}. This model describes the glucose-insulin-glucagon regulatory system by coupling ODE models for whole-body pharmacodynamics, sub-organ distribution and intracellular metabolism. The individual models describe completely different aspects, and unlike the first-principle models described in Section~2.2.2, the models do not provide different resolutions (e.g., micro- and macro-scale). Even the same class of mathematical models, namely ODEs, has been used to capture the different processes.

Multi-physics modelling requires the coupling of different models. This can be challenging, even if the individual models describe similar processes and evolve on similar time- and length-scales~\cite{NealCoo2014,KrauseUhl2010}. Beyond interface variables, which are discussed in Section~2.2.4, tailored simulation approaches have to be introduced. Walpole et al.~\cite{WalpolePap2013} introduced the terms \textit{series simulation}, \textit{parallel simulation} and \textit{integrated simulation} to provide a nomenclature for categorising simulation approaches. Series simulation acquire information by simulating one scale and provide this information as input to the next level (without feedback). Parallel and integrated simulation require communication (of different intensities) between simulations.

Parallel and integrated simulations are often computationally demanding. To circumvent this problem, approximations based on \textit{spatial homogenisation} and \textit{time-scale separation} have been introduced. The underlying assumption of spatial homogenisation is that in a particular volume element the process with the large length scale is roughly constant. As the process on the large length scale is constant over this volume element, a single evaluation of the small length scale process could be used to analyse the respective volume~\cite{ChapmanShi2008}. On the tissue level, the homogenisation would basically assume that the concentration of a substance in a particular region of an organ is constant.  Although this region might be occupied by many cells, as all cells perceive the same or at least similar environment, it is appropriate to assume that all cells behave similarly and merely a representative cell is evaluated. Using this and related homogenisation approaches, the effective resolution required on the smaller scales can be reduced, which decreases the overall complexity of the simulation~\cite{ErbertsederRei2012,RoehrleSpr2013}. An alternative to spatial homogenisation is provided by the \textit{adaptive tabulation multi-scale approach}~\cite{Pope1997,Hedengren2008}. This method exploits that different simulations with similar input sequences and model parameters are required to reduce the computational complexity.

\begin{figure*}[t]
\centering
\includegraphics[width=\textwidth]{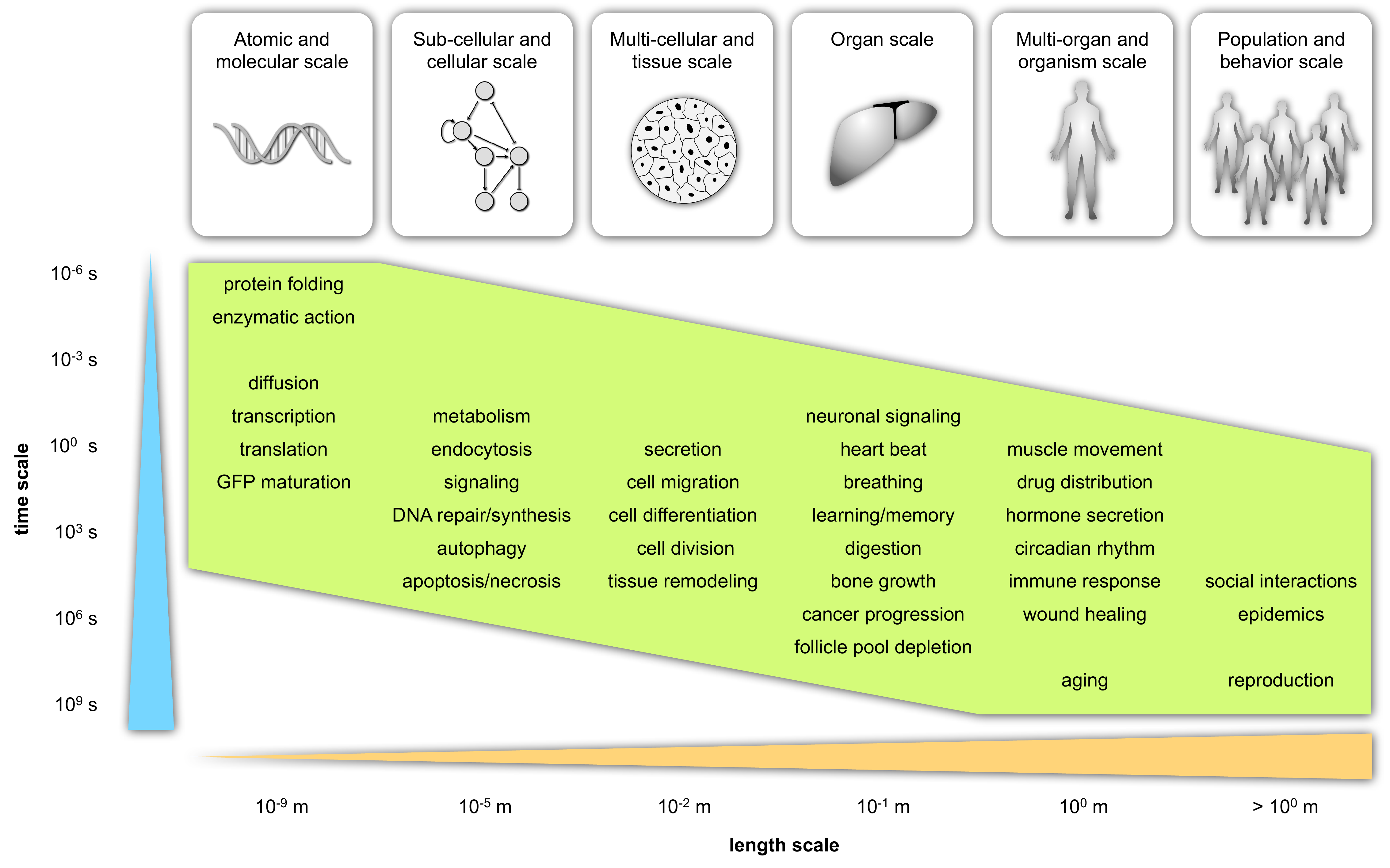}
\caption{Relation of spatial and temporal scales. For a variety of biological processes the spatial and temporal scales are roughly indicated. The list is not comprehensive, but we find a clear correlation of spatial and temporal scale.}
\label{fig: time and length scale dependence}
\end{figure*}

In addition to the spatial properties, large differences in time-scales of different processes can be used to simplify simulations. This has already been exploited by Michaelis \& Menten~\cite{MichaelisMen1913} for the study of enzymatic reactions as well as in many other fields, such as stochastic simulations~\cite{HaseltineRaw2002}. In the last decades these ideas have been extended to multi-scale processes. Multi-scale numerical schemes nowadays allow for the efficient simulation of coupled ODE-PDE systems~\cite{Whiteley2008}. Furthermore, in agent-based modelling time-resolved simulation of diffusion is circumvented based on time-scale separation~\cite{Jagiella2012,HoehmeBru2010}. 
\\[2ex]
\textbf{Remark:} Biological processes span many spatial and temporal scales. Intuitively, one might assume that microscopic processes are faster than macroscopic processes and, indeed, we find a correlation (Figure~\ref{fig: time and length scale dependence}). However, if we are interested in a particular biological process this does not have to hold. The example of anti-cancer treatment, involving drug distribution on the organism scale and apoptosis/necrosis induction on the cellular scale, shows that microscopic processes can be slower than macroscopic processes. This has to be appreciated during model and algorithm development.

\subsubsection{Coupling of models and interface variables}
\begin{figure}[t]
\centering
\includegraphics[scale=0.5]{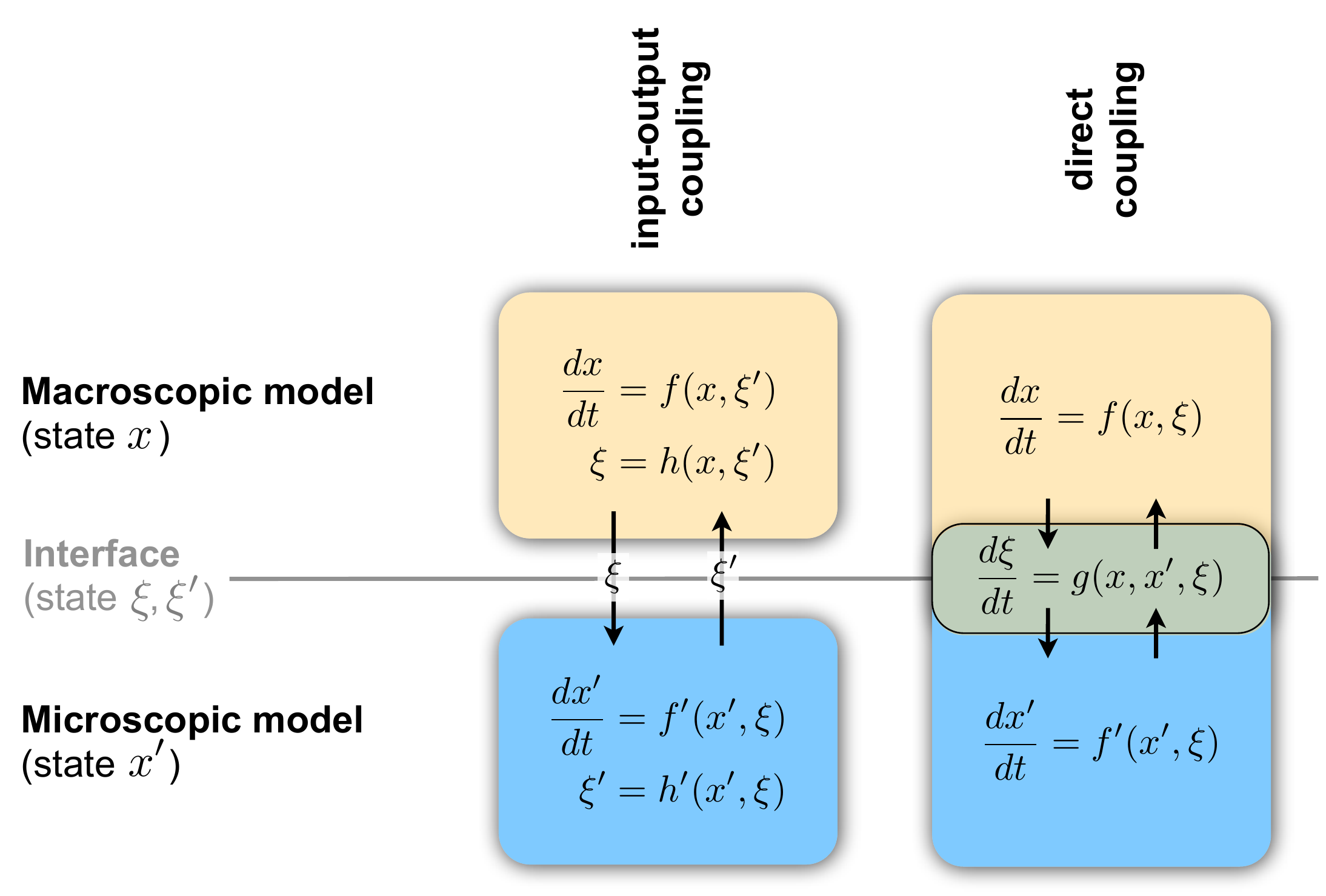}
\caption{Illustration of input-output coupling and direct coupling. Input-output coupling provides an instantaneous interaction, while in direct coupling the interface possesses dynamics and is shared between micro- and macro-scale models.}
\label{fig: interface}
\end{figure}

A key ingredient of multi-scale and multi-physics models developed to achieve Aim~1 and~2 is the coupling of models. Approaches to couple individual scales are as diverse as modelling approaches for individual biological scales. However, despite differences, an essential element of any coupling scheme is the definition of interface variables. These are properties of microscopic models which influence the macroscopic model or vice versa. We propose to distinguish two classes of interface variables and physical coupling types: \textit{input-output coupling} and \textit{direct coupling}. In input-output coupling each interface variables depends either on the state of the microscopic system or on the state of the macroscopic system~(Figure~\ref{fig: interface}, left), but not on both. The interface variables are therefore inputs and outputs for the individual scales. In direct coupling, interface variables are states shared between microscopic or macroscopic model~(Figure~\ref{fig: interface}, right). 

Examples for interface variables on the cellular and tissue scale are the instantaneous secretion rate of a molecule from a cell which enters as source term in the spatial simulation of the tissue, and the tissue level concentration of a diffusive ligand which is bound to individual cells. The instantaneous secretion rate depends only on the current cellular state. Hence, instantaneous secretion rates provide an input-output coupling. The ligand concentration is directly influenced by tissue-level diffusion and cellular binding. Accordingly, the state is shared between microscopic or macroscopic model, which establishes an overlap of the models and a direct coupling. This direct coupling is often captured by models for the interfaces.

Models only containing input-output couplings are generally easier to simulate and to analyse as modularity is ensured~\cite{Kitano2002b}. The overall problem can be decomposed in subproblems for which tailored methods might be available~\cite{EhlersZin2013}. Even toolboxes for generic (PDE) problems are available for input-output type coupling~\cite{LarsonJac2005}. 

Multi-scale models including one or more direct couplings require integrated simulation methods. Such methods have been, for instance, developed for coupled quantum mechanic and molecular mechanic simulations~\cite{WarshelLev1976}. In quantum mechanic / molecular mechanic (QM-MM) models, e.g., for enzymatic reactions, the active centres are resolved using quantum mechanic models, while the rest of the potentially large molecule is formulated using molecular mechanic models. For the combined model a joint energy function is derived which can be used for simulation.
\\[2ex]
\textbf{Remark:} The coupling type depends on the choice of interface variables. Hence, (potential) degrees of freedom can be used to render analysis and simulation more tractable.

\subsubsection{Case studies}
\begin{figure*}[t]
\centering
\includegraphics[width=\textwidth]{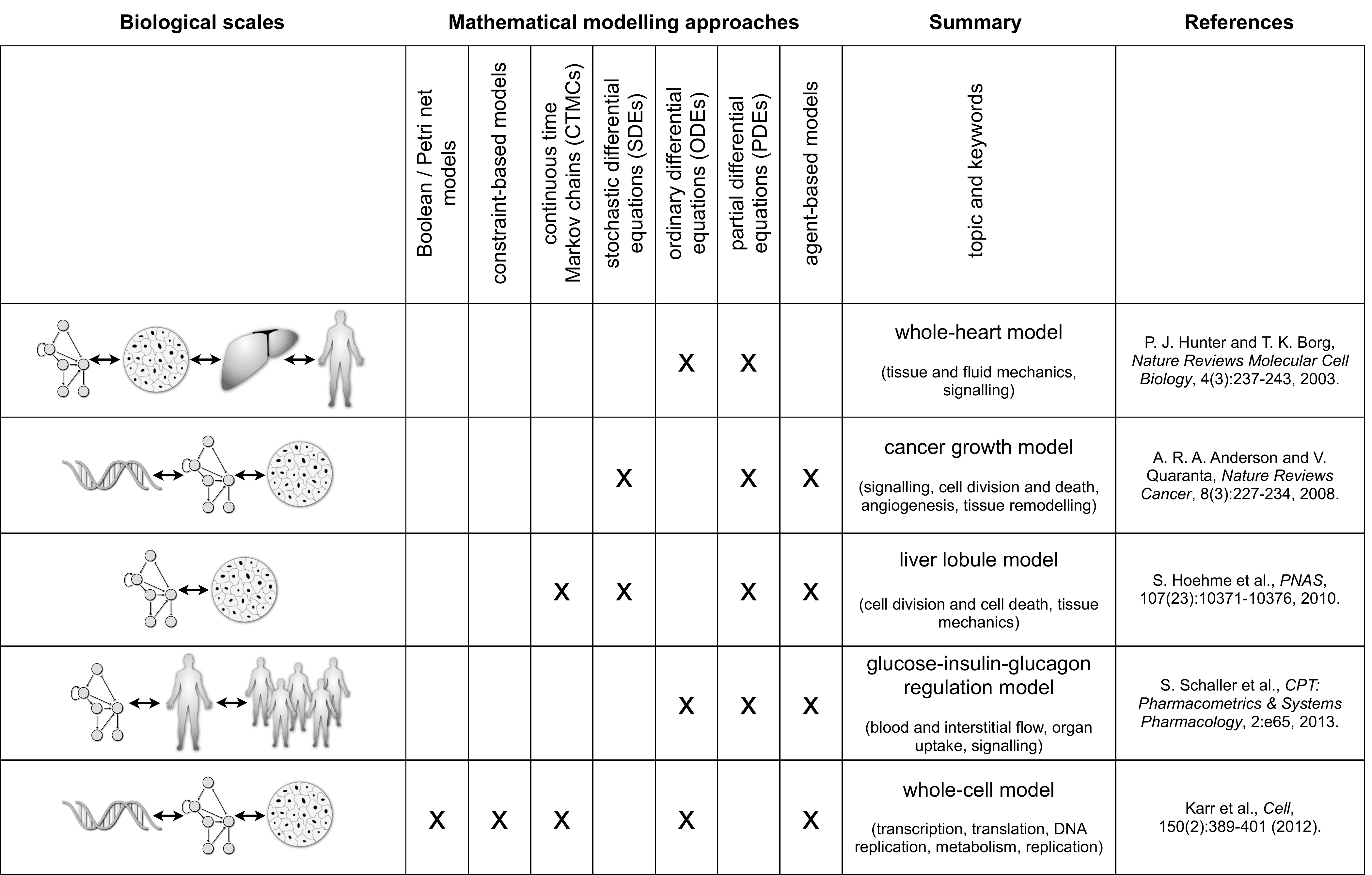}
\caption{Selection of well-known multi-scale models. Biological scales captured by the model, mathematical modelling approaches, topics and keywords, as well as a reference to an overview paper is provided.}
\label{fig: case studies}
\end{figure*}

The methods introduced in the previous sections have been used to develop multi-scale models for broad spectrum of biological processes. In this section, we will briefly present five well-known models and discuss the structure of the respective models. A visual summary of the case studies is depicted in Figure~\ref{fig: case studies}.
\\[1ex]
\textit{Whole-heart model:}
The heart was among the first organs for which holistic models were developed~\cite{HunterBor2003}. As part of the Physiome Project, bidomain models -- coupled ODE-PDE models -- for cardiac electromechanics and cardiac electrophysiology have been developed~\cite{PullanTom2002,Trayanova2011,NielsenLys2013}. Based on 3D imaging data, molecular data and electrocardiograms, these whole-heart models account for tissue deformation and heterogeneities, transport processes, as well as intracellular signalling. The different macroscopic models are mostly linked using input-output coupling and integrated simulations are employed. A transfer of these models into clinical practice, e.g., to improve diagnosis and risk assessment, seems feasible. Ischemic lesions have been detected in individual patients using model-based multi-scale data integration, and effect of drugs on pump functions can be predicted~\cite{NielsenLys2013}.
\\[1ex]
\textit{Cancer growth model:}
As cancers are among the most common causes of death, the field of integrative mathematical oncology developed. This field seeks to develop multi-scale models for cancer progression and the delineation of key cancer mechanisms~\cite{AndersonQua2008}. To achieve this aim, hybrid discrete-continuum models describing discrete individual cells and continuous distributions of nutritions and signalling molecules are used. Single-cell dynamics are governed by stochastic or deterministic models, while cell-cell interaction is captured using agent-based on-lattice/lattice-based or off-lattice/lattice-free models~\cite{Jagiella2012}. The properties of this first-principle multi-scale model are evaluated using integrated simulations.
  Parameterised multi-scale models already contributed to the understanding of complex spatial phenomena, such as fingering~\cite{AndersonQua2008}, and will probably play a key role in the development of personalised treatment strategies. Key advantages of these model-based approaches are that the impact of genetic alterations on cellular phenotypes and ultimately on the cancer progression and treatment response might be predicted, and that patient-specific data about molecule characteristics, tumor form and stage can be accounted for~\cite{HendersonOgi2014}.
\\[1ex]
\textit{Glucose-insulin-glucagon regulation model:}
Our understanding of diabetes -- one of the world's leading health problems -- has recently been improved using a multi-scale model for glucose-insulin-glucagon regulation. Using a high-dimensional system of ODEs, the pharmacokinetics and pharmacodynamics of glucose, insulin, and glucagon are modelled and coupled~\cite{SchallerWil2013}. This ODE system has been obtained by interface coupling an multi-organ model and sub cellular model. Using multi-scale experimental data, this model has been parameterised and adapted to individual patients. The resulting patient-specific models provided accurate predictions for glucose, insulin and glucagon dynamics after food intake. Model-guided development of novel diabetes treatment strategies, hence, becomes feasible.
\\[1ex]
\textit{Liver lobule model:}
A variety of models improved our understanding of tissue scale processes. Models for liver lobules, the smallest organisational units of the liver, are particularly interesting examples~\cite{HoehmeBru2010}. Using high-resolution 3D imaging, realistic hybrid discrete-continuum approaches (see explanation above) of liver lobules have been developed. These 3D models facilitated an accurate description of toxification and repopulation processes. Predictions about the cell alignment made using these models could be confirmed experimentally and enabled and improved understanding of cell-cell interaction.
\\[1ex]
\textit{Whole-cell model:}
A groundbreaking contribution to cell biology has been made by the whole-cell model for the human pathogen Mycoplasma genitalium~\cite{KarrSan2012}. This model captures all essential cellular processes, including gene regulation, metabolism, signalling and cell division. The individual processes are described using Boolean and probabilistic models, constraint-based models and ODEs. For numerical simulation an integrated simulation approach has been employed, which enables the assessment of single-cell and population dynamics. As a key feature the whole-cell model could predict transciptomics, proteomics and metabolomics data as well as the effect of gene knockouts with a surprisingly high accuracy~\cite{KarrSan2012}. This confirmed that large-scale models derived from already established multi-scale data sources can facilitate biological discovery. 
\\[3ex]
In summary, biological processes are versatile and so are modelling approaches. There is no right or wrong, but it depends on the purpose of the model (and the personal preferences of the modeller). Multi-scale models are obtained by coupling models for different processes hierarchically or in parallel. The resulting models provide first-principle formulations for macro-scale processes or simply account for different physical processes. In both cases, a more holistic description of biological processes is achieved. In particular if its mathematical description is rigorously analysed and is parameters reliable, novel insights can be gained. The rigorous analysis of multi-scale models as well as the estimation of their parameters from experimental data requires sophisticated methods.

\section{Analysis and simulation}
\label{sec: analysis and simulation}
\begin{figure*}[t]
\centering
\includegraphics[width=\textwidth]{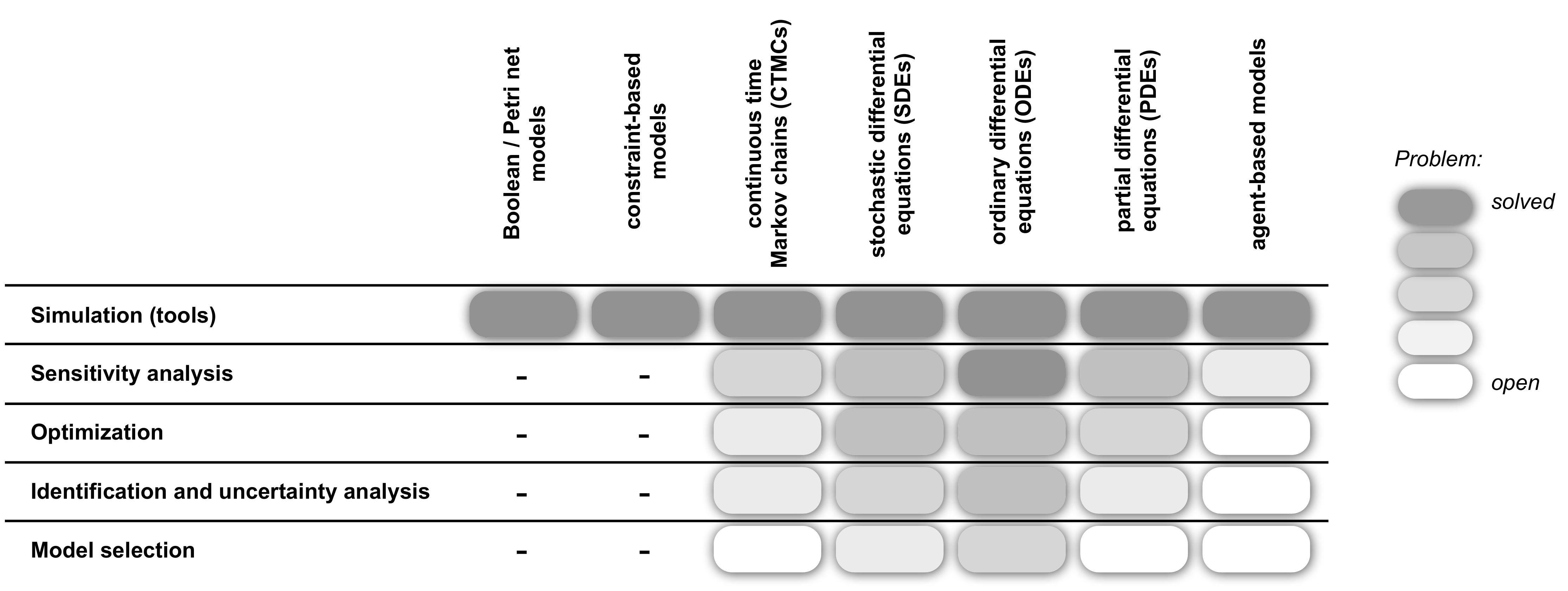}
\caption{Visual summary of the development status of methods available for different model classes. The transparency indicates the degree to which the different tools and methods are established, respectively, for which model size they are applicable. Problems for which scalable, standardised approaches are available are considered to be ``solved''. The colouring indicates trends but a quantitative interpretation is not intended. For a detailed discussion see Section~3 and~4.1.
}
\label{fig: methods}
\end{figure*}

The mathematical modelling of biological processes requires (and facilitates) a rigorous formulation of hypotheses, the first steps towards mechanistic understanding. The derived models are, in general, studied using analytical and numerical methods to gain deeper insights. Of particular interest are often
\begin{itemize}
\setlength{\itemsep}{-1mm}
\item existence and uniqueness of solutions,
\item sensitivity of solutions with respect to parameters,
\item local and global stability of steady states,
\item bifurcation structures of steady states, and
\item dynamic properties.
\end{itemize}
The analysis of these properties can already for individual model classes be demanding. While standard simulation frameworks for most model classes are available~\cite{SchaffFin1997,StilesBar2001,HoopsSah2006,KlamtSae2007,EissingKue2011}, for stochastic models even important tasks, such as sensitivity analysis~\cite{PlyasunovArk2006,KomorowskiCos2011}, pose challenges. This limits the study of the models and might explain the frequent usage of ODE and PDE models, for which a broad spectrum of methods is available, across spatial and temporal scales (Figure~\ref{fig: models for individual scales}). An illustration of the availability and sophistication of methods for different model classes is provided by Figure~\ref{fig: methods}. 

In Section~2.2.5, we reviewed multi-scale modelling approaches and presented a few well-known application examples. Interestingly, a detailed evaluation of these examples revealed that for most multi-scale systems merely simulation studies are performed. A coherent theory, in particular for coupled multi-physics models, is often missing. Methods to study existence and uniqueness, sensitivity, local and global stability and bifurcation structures are urgently needed to gain a more profound understanding of multi-scale processes. Fortunately, some brave researchers work on such methods. Among others, new theoretical methods for existence and uniqueness analysis~\cite{SonnerEfe2011,KelkelSur2012,EfendievBook2013} and stability analysis~\cite{Efendiev2010} for multi-scale ODE-PDE models have been established. However, despite significant progress, the spectrum of models which can be analysed rigorously is still rather limited. This might be related to the following three challenges:
\\[1ex]
\textbf{Challenge\;1.1:}\;\textit{The applicability of established methods has not been extended to coupled multi-scale models.}
\\[1ex]
The list of methods for which this holds is extensive. Prominent examples are methods which exploit properties of the individual (decoupled) models to study the properties of the coupled system, e.g. the small-gain theorem~\cite{KhalilBook2002}. In systems and control theory the small-gain theorem is used to prove stability of coupled ODE models. Although the underlying idea is simple, no general extensions to coupled ODE-PDE systems are available.
\\[1ex]
\textbf{Challenge\;1.2:}\;\textit{Directions from which the analysis of multi-scale models should be approached are often not obvious.}
\\[1ex]
Agent-based models for tumour growth~\cite{Jagiella2012} or tissue remodelling~\cite{HoehmeBru2010} for instance can either be interpreted as stochastic PDEs with complex noise term or as coupled Markov jump processes. These different perspectives require different mathematical and computational tools. For a satisfactory analysis probably combinations of the existing tools and theoretical results are necessary and have to be established.
\\[1ex]
\textbf{Challenge\;1.3:}\;\textit{Classical concepts and tools are often not applicable to novel classes of multi-scale models.}
\\[1ex]
Properties can easily loose their meaning if alternative model classes are considered. While stability is doubtless one of the most important properties of deterministic systems, even for simple stochastic processes stability is not easily define~\cite{Socha1986}. For stochastic systems properties such as uni- or multi-modality of the instantaneous probability density~\cite{ThomasPop2014} or statistics of noise-induced oscillation~\cite{LangWal2009} are often more essential. As models and model classes change, the properties of interest have to be adapted and generalised.
\\[2ex]
Despite the multilayer nature of these challenges, there is reason for hope. Walpole et al.~\cite{WalpolePap2013} outlined in a recent review that most multi-scale models capture only a small subset of the scales. Accordingly, not all combinations of model classes have to be considered, but already a few combinations might be sufficient. As coupled ODE-PDE systems and coupled CTMCs-PDE systems are widely used, these classes of multi-scale models might be appropriate starting points.

In addition to analytical results, equation-free stability, sensitivity and bifurcation analysis methods have been developed~\cite{SchroffKel1993,GearKev2002}. These methods allow for the simulation-based analysis of quantitative and qualitative properties of multi-scale methods. While the use of simulations allows for a certain independence from the underlying equations, the detailed mathematical understanding is lost. Furthermore, instead of analysing the mathematical model, often the numerical implementation is studied. To circumvent this, the simulation uncertainty can be quantified and taken into account~\cite{IntervalAnalysis,ChkrebtiiCam2014}.
\\[2ex]
In summary, the scarcity of the available methods limits the analysis of multi-scale models. While we cannot provide a comprehensive overview here, the need for the development of novel tools is apparent. One possibility might be to exploit modularity and merely rely on the input-output characteristics of individual models. Although potentially conservative, methods relying on this idea might be widely applicable.

\section{Parameter estimation and model selection}
\label{sec: parameter estimation and model selection}
Mathematical modelling allows for the description of biological processes and the study / prediction of their properties. However, reliable quantitative results require accurate model structures and model parameters (e.g., reaction rates and diffusion coefficients). Due to experimental constraints, structures and parameters of biological systems can often not be assessed directly. Instead, they have to be inferred from limited, noise-corrupted data. Therefore, parameter estimation and model selection methods are required.

Parameter estimation and model selection are means to integrate data from different data sources. This becomes increasingly important as the availability and variety of data increases continuously, due to new technologies. It has been shown that already model-based data integration on individual scale can significantly improve our biological understanding. As multi-scale models are significantly more complex, a further improvement has to be expected. Furthermore, information on one scale can help to understand other scales.

\subsection{Challenges for data-driven multi-scale modelling}
\label{sec: Challenges for data-driven multi-scale modelling}
Multi-scale models are obtained by coupling models for individual biological scale (Section~2.2.2 and~2.2.3). Accordingly, the most naive approach is to perform parameter estimation and model selection for the individual scales separately. As parameter estimation and model selection are fields of active research since decades (see~\cite{Tarantola2005} and references therein), a number of breakthroughs have been made for individual model classes:
\begin{itemize}
\item \textbf{Boolean and constraint-based models:} 
Boolean and constraint-based models are parameter-free and neither parameter estimation nor uncertainty analysis methods are required. For the reconstruction of Boolean models from experimental data optimisation-based methods are commonly used. Their computational complexity scales reasonably with the network size and they are applicable to networks of several hundred components~\cite{Porreca2010}. Constraint-based models are mostly derived from (literature) knowledge about the biochemistry of the underlying process. If sufficient data are available also statistical approaches, such as  Gaussian Graphical Models, can be used~\cite{TohHor2002,KrumsiekSuh2011}.
\item \textbf{ODEs:}
For ODE models robust and scalable inference methods are available~\cite{BockPli1984,Banga2008,Engl2009}. Optimisation~\cite{RaueSch2013} and uncertainty analysis~\cite{HugRau2013} of ODE models with a few hundred parameters is feasible on standard computer infrastructures. The methods are established for a wide range of applications and implemented in several software packages (see~\cite{RaueSch2013,SchmidtJir2006} and references therein). The results allow for model selection and optimal experimental design. Model selection can become time consuming if many model alternatives have to be tested, and the genom-scale models with thousands of state variables and parameters are currently beyond reach~\cite{StanfordLub2013,HendersonOgi2014}.
\item \textbf{PDEs:} 
Inference for PDE models~\cite{Banks1989,OptimizationwithPDEconstraints} is generally more challenging than for ODE models. Nonetheless, rigorous optimisation and uncertainty analysis of low- and medium-dimensional PDE systems with dozens of parameters is feasible~\cite{MuellerTim2004,LuzyaninaRoo2009,MenshykauGer2013,HockHas2013}. Based on existing simulation environments~\cite{LoggMar2012}, toolboxes for PDE-constrained optimisation are being introduced~\cite{FarrellHam2013} and are expected to replace problem-specific implementations. 
\item \textbf{SDEs and CTMCs:} 
Standard Bayesian and frequentist approaches to parameter estimation consider the conditional probability of the data given the parameters, commonly known as the likelihood~\cite{Tarantola2005}. For stochastic models, such as CTMCs and SDEs, evaluating the likelihood requires marginalisation over all possible paths of the stochastic process. This is often computationally intractable and limits the applicability of optimisation methods. Approximations to likelihood functions~\cite{Fuchs2010} as well as likelihood-free estimation methods have been introduced~\cite{MarjoramMol2003,SissonFan2007}. In particular, Approximate Bayesian Computing (ABC) methods are nowadays widely used~\cite{ToniWel2009}. These flexible methods enable the sampling of the posterior distributions, uncertainty and model selection, without evaluation of the likelihood function. Due to the large number of simulations, mosty problems with rather low and medium computational complexity have so far been studied~\cite{ToniWel2009,LiepeBar2010,LillacciKha2013}. For certain subclasses of models also software packages are available~\cite{LiepeBar2010}.
\item \textbf{Agent-based models:} 
In agent-based modelling often interacting and non-interacting agents are distinguished. Models consisting of non-interacting agents are widely used in pharmacokinetics and pharmacodynamics, and are -- as mentioned above -- also denoted as mixed-effect models. For mixed-effect models standard inference methods are available~\cite{Pinheiro1994,ZechnerPel2012} and implemented in free and commercially available software packages~\cite{TornoeAge2004}.\\
Models of interacting agents are more difficult to parameterise than models of non-interacting agents. The reason is the need for integrated simulations of the complete system. The resulting computational complexity is challenging and mainly problem-specific solutions are currently employed~\cite{ThorneHay2011}. One reason is the variety of experimental data and their corresponding objective function types.
\end{itemize}

The integration of data-driven models for individual scales to multi-scale models is reasonable and rather tractable.  Therefore, this approach has been successfully applied in the literature several times~\cite{tenTusscherNob2004,HayengaTho2011}. However, this approach suffers also from several limitations. It is clear that interface variables have to be consistent across scales. This cannot be guaranteed if model fitting is done separately. Furthermore, for multi-scale models with direct coupling a separate data-driven modelling might not even be feasible. Beyond these technical aspects, experimental data used for the data-driven modelling of the individual scales might have been collected under different conditions. Accordingly, the models might not be valid for the regimes required in a multi-scale model. Extensive studies of the transferability of models are currently not available.

The aforementioned limitations establish the need for an integrated approach for data-driven multi-scale modelling. Beyond the data-driven modelling of individual scales, an important first step (if feasible), the multi-scale model has to be fitted to / validated with multi-scale data~\cite{WalpolePap2013}. Examples for this have been provided in the context of blood cell mechanics~\cite{FedosovLei2011} and heart electromechanics~\cite{NielsenLys2013}.

Parameter estimation, uncertainty analysis and model selection for the multi-scale models is apparently more challenging than data-driven modelling of the individual scales. The dimension of the parameter space increases and the handling of the models becomes more demanding. Key challenges are in particular: 
\\[1ex]
\textbf{Challenge\;2.1:}\;\textit{Computational complexity for multi-scale simulation (and sensitivity analysis) is often large.}
\\[1ex]
\textbf{Challenge\;2.2:}\;\textit{Stochasticity of many multi-scale models establishes the need for a large number of simulations.}
\\[1ex]
\textbf{Challenge\;2.3:}\;\textit{Reproducibility of parameter estimation and model selection results have to be ensured.}
\\[1ex]
\textbf{Challenge\;2.4:}\;\textit{Lack of tailored and generic computational tools for data-driven multi-scale modelling.}
\\[1ex]
Some of these challenges are inherited from models of particular scales.
 
In the following, we will discuss novel approaches and ideas which could in the future be used to address these challenges. Thereby, we will provide a comprehensive list of potential methods helpful for data-driven multi-scale models.

\subsection{Optimisation}
To determine unknown model parameters, statistically motivated measures for the goodness of fit are used. These objective functions are optimised with respect to the parameters. For dynamical systems, the respective optimisation problems are in general nonlinear and non-convex. Hence, sophisticated optimisation schemes are required. Common tools employ multi-start local optimisation~\cite{RaueSch2013}, multiple shooting~\cite{Biegler2007,BockPli1984}, evolutionary algorithms~\cite{Back1996,Balsa-Canto2008c}, pattern search~\cite{Vaz2007} and particle swarm optimisation~\cite{Vaz2007,Yang2010}. Banga~\cite{Banga2008} and Weise~\cite{Weise2009} provide comprehensive surveys of local and global optimisation procedures.

In the course of the optimisation, the goodness of fit is assessed at different points in parameter space. This requires the simulation of the model and -- depending on the optimisation method -- the evaluation of sensitivities (derivatives of the model output with respect to the parameters). The repeated simulation and sensitivity evaluation can be time consuming (Challenge~2.1), stochasticity (Challenge~2.2) and reproducibility (Challenge~2.3) are however even more intricate.

Parameter optimisation for a number of multi-scale models has been approached in recent years. In particular for PK/PD models great successes have been reported (see, e.g.,~\cite{SchallerWil2013}). Here, different ODE models are coupled and simulation as well as parameter estimation remains efficient. The optimisation-based integration of experimental data collected on different scales provided novel insights in diseases, such as diabetes. Promising results have been obtained for mixed-effect models describing population and single-cell level~\cite{KallenbergerBea2014}, but reproducibility is often an issue. For coupled PDE models novel optimisation methods resulted in great successes, e.g., in the context of the Virtual Heart~\cite{NielsenLys2013}.

For models composed of different model types, optimisation seems more challenging. This is also indicated by the aforediscussed contributions (Section~2.2.5). Hoehme et al.~\cite{HoehmeBru2010} introduced and validated a sophisticated agent-based model for liver regeneration. The available measurement data were however merely used to determine realistic simulation domains and rough parameter values. A parameter optimisation was not approached. The same is true for the whole-cell model developed by Karr et al.~\cite{KarrSan2012}. Reasons for this are stochasticity and computational complexity of model simulations.

\begin{figure*}[t]
\centering
\includegraphics[width=\textwidth]{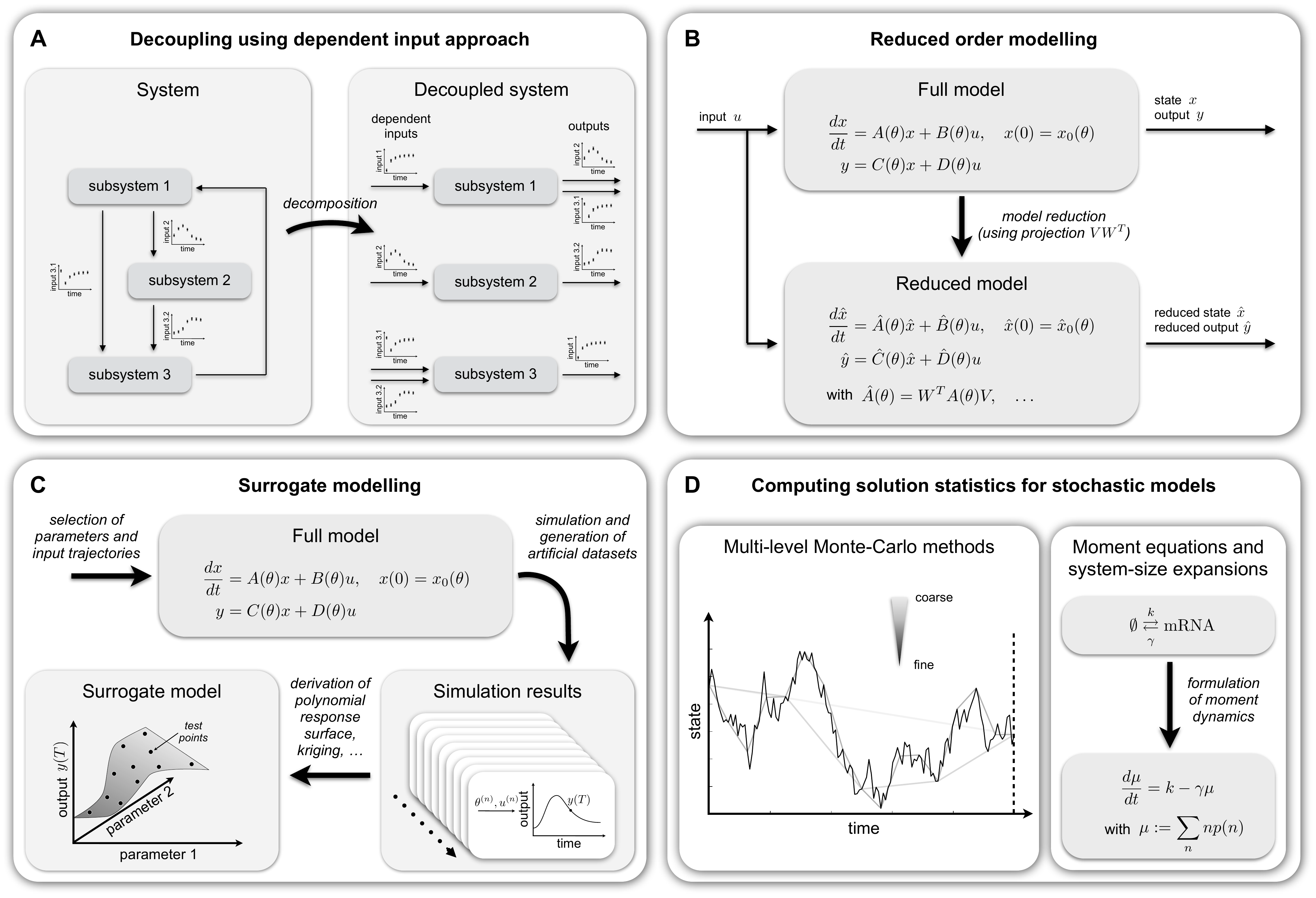}
\caption{
Visual summary of methods which could potentially improve optimisation of multi-scale models: 
\textbf{(A)} decoupling using dependent input approach;
\textbf{(B)} reduced order modelling;
\textbf{(C)} surrogate modelling; and
\textbf{(D)} computing solution statistics for stochastic processes using multi-level Monte-Carlo methods \& moment equations and system-size expansion).
}
\label{fig: optimisation methods}
\end{figure*}

There are a series of promising mathematical methods which tackle Challenges 2.1 - 2.3. These methods split the underlying optimization problem in smaller subproblems or decrease the computational complexity associated to model evaluations. We will shortly outline the key ideas underlying these methods. A visual summary is depicted in Figure~\ref{fig: optimisation methods}. 

\subsubsection{Decoupling using dependent input approach}
The parameter estimation for models with high-dimensional parameter and state space is often challenging. An intuitive idea is therefore to exploit the modularity of biological systems. The overall system can be decomposed into a set of interconnected subsystems~\cite{EdererSau2003}. The subsystems, for instance, describe individual biological scales or span different scales. A subsystems is interfaced with other subsystems. The \textit{dependent input approach} regards these other subsystems as unmodelled dynamics and replaces them by fictitious `dependent inputs'~\cite{vanRielSon2006}. This enables the decoupling of subsystems, assuming that the dependent inputs can be measured in experiments. The subsystems can then be fitted separately, assuming that they do not share parameters. This eliminates the need for multi-scale simulations and addresses Challenge~2.1. Furthermore, it provides a tailored tool for data-driven multi-scale modelling (Challenge~2.4).

The dependent input approach is rather flexible and in principle not limited to a particular class of models. It however requires that the dependent inputs are measured. This limits the decomposition and renders it also measurement dependent. Furthermore, subsystems require continuous input signals while measurement data are mostly collected at discrete time points. Interpolation and filtering approaches can be used to close these gaps~\cite{GeorgoulasCla2012}, this can however be error-prone. More sophisticated approaches fit input data and subsystems dynamics simultaneously~\cite{KaschekTim2012,SchelkerRau2012}.

The dependent input approach is widely used for optimisation~\cite{KaschekTim2012,SchelkerRau2012} as well as uncertainty analysis~\cite{WaldherrHas2011}. Open questions are mainly related to the securing of consistency between input and output signals of individual subsystems.

\subsubsection{Reduced order modelling}
The computational effort associated with numerical simulations is frequently the bottleneck for optimisation. Model order reduction methods reduce the complexity of high-dimensional model, while preserving their input-output behaviour as much as possible~\cite{Schilders2008}. The resulting reduced models, which can be simulated more efficiently, mimic the behaviour of the full model.

For linear models efficient and reliable model order reduction methods are available~\cite{Antoulas2005book}, for instance, SVD-based~\cite{Sirovich1987} and Krylov subspace methods~\cite{Grimme1997}. In the last decade these methods have been extended to linear~\cite{HaasdonkOhl2008,HaasdonkOhl2011,BaurBen2009,RozzaHuy2008} and non-linear~\cite{Lall2002,WirtzHaa2011} parametric (partial) differential equations. This enabled the use of reduced order models in optimisation~\cite{Benner2009}. Simulations of the full model are simply substituted by simulations of the reduced order model. To account for the approximation error, a-posteriori error bounds can be used~\cite{Benner2009,HasenauerLoh2012,DihlmannHaa2013}. 

In multi-scale modelling, reduced order models are easily used to decrease the computational complexity of models for individual scales or processes. Furthermore, promising multi-scale model reduction methods have been developed which consider several scales simultaneously~\cite{YvonnetHe2007,GeersKou2010}. Using these approaches a speed-up of the numerical calculations by several orders of magnitude has been achieved, rendering parameter estimations feasible by addressing Challenge~2.1. 

\subsubsection{Surrogate modelling}
Model order reduction methods exploit the structure of the governing equation. This limits the applicability of these methods -- as the equations might become high-dimensional or highly non-linear -- and led to the development of surrogate modelling approaches. Surrogate models, also known as metamodels, response surface models and emulators, are scalable analytical models for the approximation of the multivariate input-output behaviour of complex systems. Surrogate models are derived from simulated input-output data and do not consider the internal structure of the original model. Popular surrogate models are polynomial response surfaces~\cite{BoxDra2007}, Kriging~\cite{Wilkinson2011}, space mapping~\cite{BandlerDak2004}, support vector machines and radial basis function approximations~\cite{RegisShoe2007}. Similar to reduced order modelling, surrogate modelling can be used to reduce the computational complexity of a single model evaluation (Challenge~2.1).

Surrogate models have been used to approximate objective functions~\cite{RegisShoe2007} as well as time-courses of (multi-scale) models~\cite{WirtzKar2015,BandlerDak2004,Wilkinson2011}. Both types of surrogates have been used in surrogate-based optimisation. Surrogate-based optimisation methods circumvent evaluation of the computationally demanding model by evaluating the surrogate model. Based on a first space filling sampling, e.g., latin hypercube sampling, an initial surrogate model is derived (Step~1). This surrogate model is optimised (Step~2). At the new optimal point the full model is evaluated (Step~3) and the surrogate model is updated (Step~4) before returning to Step~2. The individual steps in surrogate-based optimisation are involved and there exist a variety of different approaches. For details we refer to~\cite{RegisShoe2007,MullerSho2014} and references therein.

\subsubsection{Multi-level Monte-Carlo methods} 
For stochastic models a single evaluation of the objective function requires the averaging over stochastic simulations. This averaging can be computationally demanding because it normally requires many simulations. To decrease the necessary number of simulations and to address Challenge~2.2, multi-level Monte-Carlo methods have been developed~\cite{Heinrich2001}. Multi-level methods employ series of increasingly complex models. For SDEs, this series of models could be, for instance, different numerical SDE solvers with increasing accuracy~\cite{Giles2008}. Instead of estimating the mean behaviour of the full model using Monte-Carlo integration, e.g., a numerical SDE solver with a very high accuracy, the difference of adjacent models in the series is assessed. As this difference is smaller, fewer evaluations are necessary, often resulting in a speed up and a lowered variance of estimates~\cite{Heinrich2001}. Similar approaches have recently been introduced for CTMCs~\cite{AndersonHig2012}.

Multi-level Monte-Carlo methods often use approximate stochastic simulations. These methods on their own can already accelerate simulations. The simulation of CTMCs is, for instance, often approximated using tau-leaping~\cite{Gillespie2001}, time-scale separation~\cite{HaseltineRaw2005} or diffusion approximation~\cite{Fuchs2010}.

Monte-Carlo integration induces stochasticity of the objective function. Two evaluations of the objective function for the same parameter will provide slightly different results. This is a severe issue for the commonly used deterministic optimisers, and therefore stochastic optimisation procedures have to be employed to ensure robustness~\cite{Weise2009}, e.g., implicit filtering~\cite{Kelley2011}. Such stochastic optimisers use involved update schemes, and tailoring of these schemes to the structure of the considered multi-scale model might be beneficial. In particular, block updates might improve the search performance~\cite{WilkinsonYeu2002}. Block update schemes, currently used in Markov-chain Monte-Carlo methods, can exploit the model structure to virtually reduce the problem size by updating merely parameters belonging to the same scale or the same process.
 
\subsubsection{Moment equations and system-size expansions}
Alternatives to Monte-Carlo integrations are provided by moment equations~\cite{Engblom2006} and system-size expansions~\cite{Grima2010}. Moment equations and system-size expansions are deterministic models describing the statistics -- mean, variance and higher-order moments -- of the solutions of stochastic processes, i.e., CTMCs, which eliminates the need for repeated stochastic simulations (Challenge~2.2). A further advantages is that moment equations and system-size expansions are deterministic models which allow for the use efficient deterministic optimisers. For these optimizers the reproducibility is general good, addressing Challenge~2.3. The key disadvantage of moment equations and system-size expansions is that they merely provide approximations, as moment closure and/or truncations of infinite series are required~\cite{Gillespie2009}. Detailed studies of the influence of the approximation error on the parameter estimation are so far missing. To minimise the effects, hybrid methods have been developed, which use a fully stochastic description for low-copy number species and a moment-based description for high-copy number species~\cite{MenzLat2011,Jahnke2011,HasenauerWol2014,ThomasPop2014}. The resulting hybrid models are often more accurate, but their simulation is computationally also more demanding.

Moment equations are available for concentrated as well as distributed processes. Already in the 1990s, spatial moment equations have been developed for individual-based models~\cite{BolkerPac1997,LawDie2000}. These equations have been successfully used to study population dynamics on regular lattices, irregular networks, and continuous spatial domains. Different closure schemes have been developed to provide appropriate approximation accuracies~\cite{GandhiLev2000}. The consideration of long-range correlations, which improves the approximation accuracy but also increase the computation time, turned out to be particular critical.
\\[3ex]

\subsection{Identifiability and uncertainty analysis}
As experimental data are limited and noise-corrupted, parameter estimation has to be complemented by structural and practical identifiability. Structural identifiability is concerned with the structure of the system~\cite{CobelliDiS1980} and can be used to assess a priori whether the parameters can in principle be determined from noise-free data. In contrast, practical identifiability is related to a particular dataset and its information content~\cite{Raue2009}. Using practical identifiability analysis the parameter and prediction uncertainties are quantified in terms of asymptotic and finite-sample confidence intervals.

Asymptotic confidence intervals can be computed via local sensitivity-based methods, e.g., the Wald approximation~\cite{Meeker1995} or the Fisher information matrix (FIM)~\cite{MurphyVaa2000}. While these asymptotic confidence intervals are already determined by standard optimisers, these confidence intervals are not reliable. Therefore, finite sample confidence intervals derived using bootstrapping~\cite{Joshi2006}, profile likelihoods~\cite{Raue2009} and Markov chain Monte-Carlo methods~\cite{Wilkinson2007} are preferred in practice. A recent study showed however, that also the use of bootstrapping-based confidence intervals are problematic~\cite{FroehlichThe2014}. While profile likelihoods and bootstrapping require a large number of local or global optimisations, Bayesian methods exploit sampling of posterior distributions using, e.g., Markov chain Monte-Carlo methods. For multi-scale models using a coherent modelling of the scales, this is often feasible using current methods~\cite{SchallerWil2013}, while for most multi-scale models this is so far intractable.

These limitations can be partially overcome using aforementioned reduced order and surrogate models, as well as, multi-level Monte-Carlo methods and moments equations. In addition, tailored and efficient identifiability and uncertainty analysis methods are being developed. A few of those will be introduced in the following.

\subsubsection{Structural identifiability analysis for interconnected systems}
Identifiability analysis uses the governing equations of dynamical systems to study whether parameters can in principle be determined~\cite{CobelliDiS1980}. Therefore, several approaches based on Taylor and generating series, implicit function theorem and differential algebra have been developed~\cite{ChisBan2011}. These methods are applicable to ODE and PDE models. The key advantage of these methods is that they exploit symbolic calculations and do not rely on numerics. This makes them robust, but also limits their application to small- and medium-size systems. To overcome these challenges, approaches for interconnected systems have been developed~\cite{Glad2006,GerdinGla2007}. These methods exploit the modular structure of interconnected / large-scale systems and the input-output structure of subsystems to decide about global structural identifiability. Accordingly, these methods are well-suited for the study of multi-scale models and address Challenge~2.4. 

\subsubsection{Simulation-based profile likelihood calculation}
In addition to the development of novel methods designed for multi-scale models, the efficiency of existing methods is continuously increasing. This allows for a more rigorous assessment of problems at hand and improves the reproducibility of results (Challenges~2.3). Among others, fast simulation-based profile likelihood calculations methods have been proposed~\cite{ChenJen2002}. These methods circumvent the repeated local optimization by formulating dynamical systems evolving along the profiles. Using gradient and hessian of the objective function, the update direction for the parameters is determined directly. The resulting path in parameter space describes the profile likelihoods. First (unpublished) results indicate a significant acceleration.

\subsubsection{Efficient sampling methods}
To improve the efficiency of Bayesian uncertainty analysis, structure exploiting sampling schemes have been introduced. In particular, Riemann manifold and Hamiltonian Monte-Carlo methods provide a performance boost for a variety of applications~\cite{Neal2011,GirolamiCal2011}. These methods use gradient and hessian of the objective function to construct a local approximation. By sampling from this local approximation, a significantly reduction of the autocorrelation of the Markov chain samples can be achieved compared to standard Markov chain Monte-Carlo methods. Similar efficiency increases could also be achieved using adaptive single- and multi-chain Monte-Carlo methods~\cite{HaarioLai2006,MiasojedowMou2012,HugRau2013}. These adaptive samplers use the available sample path to construct an approximation of the local structure. Accordingly, the potentially demanding evaluation of gradient and hessian is circumvented.

Samplers can also be combined with surrogates, e.g., radial basis function approximation of the posterior distribution~\cite{FroehlichHro2014}. Furthermore, surrogate-based sampling schemes have been developed~\cite{HigdonRee2011}. For computationally intensive problems these approaches can outperform standard methods. This natural extension of surrogate-based optimisations should be explored in the future, in particular, as information collected during optimisation can be reused.

\subsubsection{Approximate Bayesian Computing}
The evaluation of likelihood functions, the statistical distance of model and data, is for many models difficult. Therefore, likelihood-free methods have been proposed~\cite{MarjoramMol2003}, also known as Approximate Bayesian Computing (ABC) methods. ABC methods circumvent the evaluation of the likelihood function using model simulations~\cite{ToniWel2009} and thereby address Challenge~2.2. A simulation for a particular parameter value is accepted or rejected based on simple distance measures. For appropriate choices of the distance measure, ABC methods sample approximately from the Bayesian posterior distribution. These samples enable a direct assessment of parameter and predictions uncertainties. 

To ensure efficiency and reliability of ABC methods, a variety of sophisticated sampling schemes have been developed. Popular are ABC Markov chain Monte-Carlo and ABC sequential Monte-Carlo methods~\cite{ToniStu2010}. For models with time-consuming simulations also surrogate-based approaches have been developed, i.e., approximate ABC (AABC)~\cite{BuzbasRos2013}.

\subsection{Model selection}
In biology not only model parameters but also model structures are often unknown. Hence, subsequent to parameter estimation, model selection has to be performed to evaluate competing hypotheses. Therefore, a set of alternative models is defined and the best model is selected using likelihood ratio~\cite{Wilks1938}, Akaike information criterion~\cite{Akaike1973}, Bayesian information criterion~\cite{Schwarz1978}, Bayes factors~\cite{KassRaf1995} or flavours of these criteria. If the set of alternative models is too large, forward and backward selection methods can be used to avoid a full enumeration and exploration. The aim of model selection is to determine the underlying biological and biochemical mechanism and to avoid over- and underfitting.

Model selection requires parameter estimation for all model alternatives. An acceleration of parameter estimation methods, there also results in an accelerated model selection. In addition, more sophisticated model selection methods are being developed, e.g., more reliable algorithms for evaluating Bayes factors~\cite{Vyshemirsky2008}. This is also important, as detailed studies of ABC-based model selection methods revealed. For stochastic models ABC-based model selection is the method of choice~\cite{ToniStu2010}, however, if insufficient statistics (= error norms) are used, Bayes factors can select incorrect models~\cite{RobertCor2011}. Sufficiency conditions have been derived to avoid this problem~\cite{MarinPil2014}.

Case studies showed that the combination of data collected on different levels can enhance the model selection. In a recent paper, it has been shown that model selection using gene expression and metabolite data allows for an improved inference of regulation mechanism compared to the individual models/datasets~\cite{ChandrasekaranPri2013}. The reason for this improvement is that information about adjacent levels provide additional constraints for the model. Hence, model selection on individual scales has to be complemented by multi-scale model selection. In our opinion appropriate methods to tackle this problem are currently not available. Multi-scale models are mostly selected based on heuristic arguments and visual inspection.
\\[3ex]
In summary, this section provided an overview about state-of-the-art methods for data-driven multi-scale modelling. Challenges have been outlined and novel methods and ideas have been discussed. The discussed methods and ideas might revolutionise multi-scale data integration by shifting the bounds and allowing for much higher-dimensional problems. 

\section{Conclusions and outlook}
\label{sec: conclusion}

A multitude of modelling approaches is used to describe biological processes and to unravel the underlying working principles. This review provides an overview over the spectrum of different (multi-scale) modelling concepts, thereby not aiming for complete comprehensiveness. Qualitative and quantitative modelling approaches have been outlined along methods to integrate them in multi-scale models.

To analyse multi-scale models sophisticated tools are necessary, whose development poses interesting mathematical challenges. Existing theory has to be extended (Challenges 1.1) and different mathematical disciplines have to be linked (Challenges 1.2 \& 1.3). To facilitate this process and to bundle research activities, standard classes of multi-scale models should be defined. 

Beyond modelling and model analysis, we should also follow model-based multi-scale integration of experimental data as a field of active research. The development of parameter optimisation, uncertainty analysis and model selection for multi-scale models is in its early phase. Computational complexity (Challenge~2.1), stochasticity (Challenge~2.2), reproducibility (Challenge~2.3) and the lack of computational tools (Challenge~2.4) provide limitations to what is currently feasible. There are however many interesting ideas and approaches waiting to be explored. In particular, the use of approximation, reduced order models and surrogate models is promising. The derivation of appropriate surrogate models might even be automated as no detailed analysis of the governing equations is required but merely input-output data are needed. However, to use surrogate-based approaches with stochastic models additional robustness improvements are required.

In addition to the development of appropriate inference methods, it has to be decided which complexity is truly necessary. Sometimes simplistic models might be more appropriate than detailed models, although the later are more realistic. If the detailed models cannot be parameterised using available data and inference methods, it is unclear what we can truly learn using them. We share the opinions of two well known scientists: \textit{``Simplicity is the ultimate sophistication'' (Leonardo da Vinci)}, and \textit{``everything should be made as simple as possible, but no simpler'' (Albert Einstein)}. Accordingly, we might not search for the most appropriate and detailed model, but for the model with the highest reliable informativeness~\cite{ChehreghaniBus2012}. Extensions to such concepts to multi-scale dynamical models would be very interesting.

In summary, in the last decade multi-scale modelling already contributed significantly to improving our understanding of complex biological systems. Flagship projects, like the Virtual Heart and the Virtual Liver, illustrated how multi-scale modelling can be exploited. Nowadays, there is a broad spectrum of simulation environments for multi-scale modelling, however, methods to parameterise the models are mostly missing. Mathematical and computational research has to be fostered to solve this problem and initiatives, e.g., DREAM challenges, are needed to carefully evaluate the developed methods. The availability of inference tools might turn multi-scale modelling into a standard tool in biological sciences, similar to powerful new measurement devices. Paraphrasing the idea of Galileo Galilei: We should ``measure what can be measured and make the rest measurable'' using multi-scale models. 

\vspace*{2mm}
\section*{\normalsize Acknowledgments}
The authors would like to acknowledge financial support from the German Federal Ministry of Education and Research (BMBF) within the SYS-Stomach project (Grant No. 01ZX1310B), the European Union within the ERC grant ``LatentCauses'', and the Postdoctoral Fellowship Program (PFP) of the Helmholtz Zentrum M\"unchen.

\bibliographystyle{unsrt}
\bibliography{Database}

\end{document}